\documentclass[a4paper]{llncs}

\usepackage{amsmath} 
\usepackage{color} 

\usepackage{amssymb}
\usepackage{graphicx}

\usepackage{url}
\urldef{\mailsa}\path|{armano, nima.hatami}.diee.unica.it| 

\usepackage{algorithmic}
\usepackage{algorithm}

\newcommand{\highlight}[1]{ \noindent \textbf{#1} \vspace{0.3cm}}

\newcommand{\down}{\vspace{0.2cm}}
\newcommand{\up}{\vspace{-0.2cm}}

\begin{document}

\mainmatter  

\title{A Route Confidence Evaluation Method for Reliable Hierarchical Text Categorization}

\titlerunning{A Route Confidence Evaluation Method for Reliable HTC}
%


\author{Nima Hatami\inst{1}
\and Camelia Chira\inst{2}
\and Giuliano Armano\inst{3}}

\institute{BioCircuits Institute\\
University of California\\
San Diego, CA 92093-0328, USA\\ 
\and
Department of Computer Science\\
Babes-Bolyai University\\
Kogalniceanu 1, Cluj-Napoca 400084, Romania\\
\and
Department of Electrical and Electronic Engineering\\
University of Cagliari\\
Piazza D'Armi, I-09123 Cagliari, Italy\\
}

\maketitle

\abstract{Hierarchical Text Categorization (HTC) is becoming increasingly important with the rapidly growing amount of text data available in the World Wide Web.
Among the different strategies proposed to cope with HTC, the Local Classifier per Node (LCN) approach attains good performance by mirroring the underlying class hierarchy while enforcing a top-down strategy in the testing step. However, the problem of embedding hierarchical information (parent-child relationship) to improve the performance of HTC systems still remains open.
A confidence evaluation method for a selected route in the hierarchy is proposed to evaluate the reliability of the final candidate labels in an HTC system. In order to take into account the information embedded in the hierarchy, weight factors are used to take into account the importance of each level. An acceptance/rejection strategy in the top-down decision making process is proposed, which improves the overall categorization accuracy by rejecting a few percentage of samples, i.e., those with low reliability score. Experimental results on the Reuters benchmark dataset (RCV1-v2) confirm the effectiveness of the proposed method, compared to other state-of-the art HTC methods.}


\section{Introduction}

Text categorization is one of the key tasks in information retrieval and text mining. It is widely used in many intelligent systems, e.g., content-based spam filtering, e-mail categorization, web page classification and digital libraries \cite{Spam filtering} \cite{email classification} \cite{web classification}.
Due to some challenging characteristics, such as the huge number of sparse features and a typically large number of classes, text categorization attracted a lot of attention from different research fields, including machine learning, data mining and pattern recognition.

There are three main approaches to text categorization: (i) flat approaches, which totally ignore the class hierarchy, (ii) local approaches, which run a classifier only for a subset of the hierarchy, and (iii) big-bang approaches, which use a single classifier for the whole category space.
However, despite the variety of proposed methods, also depending on different types of classifiers and on feature selection/extraction algorithms, there is no clear outperforming method (see \cite{Hierarchical survey} and \cite{LSHC2010} for more information on this issue). 

The main idea of Hierarchical Text Categorization (HTC) is to take benefit of the information embedded in the hierarchical structure, with the goal of improving the classification performance. Browsing the massive amount of data represents a further motivation for using a hierarchical structure. 
Typically, categories are structured according to a top-down view, where
nodes at upper level as used to represent generic concepts while
nodes at lower levels are viewed as more specific categories.
Top-down error propagation is a major disadvantage of HTC methods, which implies that a misclassification made at upper levels cannot be recovered at lower levels. Some \emph{error correction} strategies have been proposed to minimize error propagation \cite{LSHC2010}, but their performance is still limited.

According to the survey paper of Silla and Freitas \cite{Hierarchical survey}, 
there are three kinds of local classifier methods, depending on how local information is used and on how local classifiers are built: i) local classifier per node (LCN), ii) local classifier per parent node (LCP), and iii) local classifier per level (LCL). Each of these approaches has its own drawbacks and benefits. In the first, each node and its corresponding classifier is independent from the rest of the hierarchy, thus facilitating the maintenance of the hierarchy, as (to some extent) the classifier associated to a node can be modified without manipulating the others.
While LCN methods employ a great number of classifiers, one for each node of the given hierarchy, LCP and LCL methods lie on non binary classifiers --which is a clear source of additional complexity for the underlying learning process. 

In this paper, an evaluation strategy for LCN methods is proposed,\footnote{It is worth pointing out that, although fraed for LCN, the strategy could be easily adapted to any other top-down local classifier methods.} which allows to evaluate the route of the underlying hierarchy that has been selected depending on the input in hand. The evaluation strategy returns a reliability measure, with the goal of deciding the confidence of the final label assignment.
Weight factors for each level of the hierarchy are used, with the goal of adding hierarchical information in the decision making process. Each weight factor is strictly related with the likelihood for an error to occur at the given level.
A thresholding mechanism is proposed to accept/reject the candidate label by considering the reliability score assigned to the candidate route in the hierarchy. Experimental results show a significant increase in categorization accuracy, obtained by rejecting a few percentage of the samples with low reliability score.

The remainder of this paper is divided as follows: Section 2 briefly recalls the LCN approach and its training/testing strategies. The proposed route reliability evaluation is described in Section 3. In Section 4, we validate our proposed method on three topics, industry and regions datasets of the RCV1-v2, the Reuter's text categorization test collection and discuss the results. Section 5 concludes the paper. 

\section{The Local Classifier per Node Approach}

LCN appears to be the most used and acknowledged approach in the hierarchical classification literature \cite{Hierarchical survey}. A local binary classifier runs on each node of a hierarchy except for the root node (whose typical responsibility is to dispatch the input to be classified to all its children). The hierarchical information and parent-child relationship is taken into account by defining the set of positive and negative examples while training each classifier.
The decision making process starts from the root node and proceeds downward to the lower levels of a hierarchy. Figure~1 illustrates this approach with an example. 

\begin{figure}
\centering
\includegraphics[width=5.5cm,height=4.5cm]{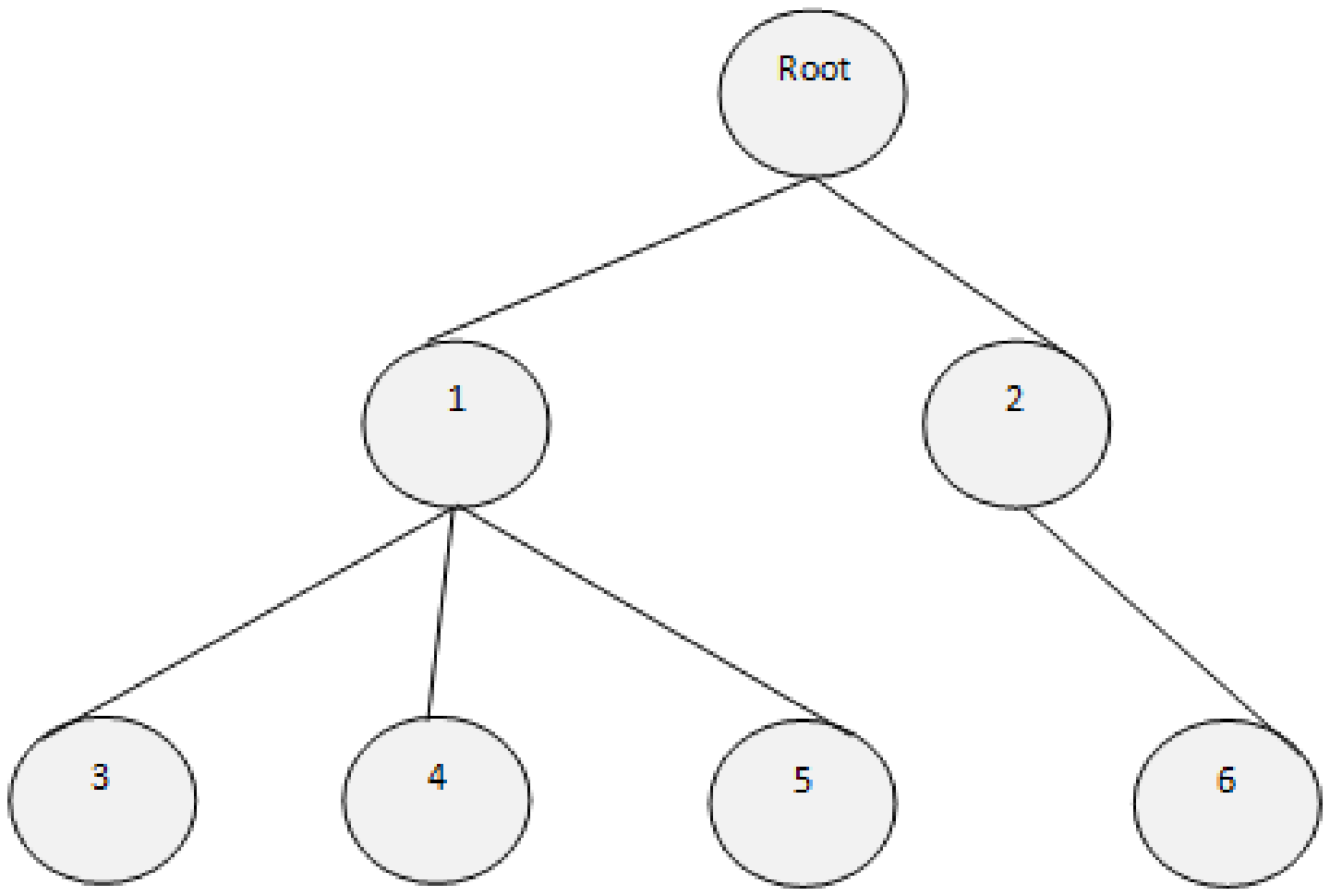}
\includegraphics[width=5.5cm,height=4.5cm]{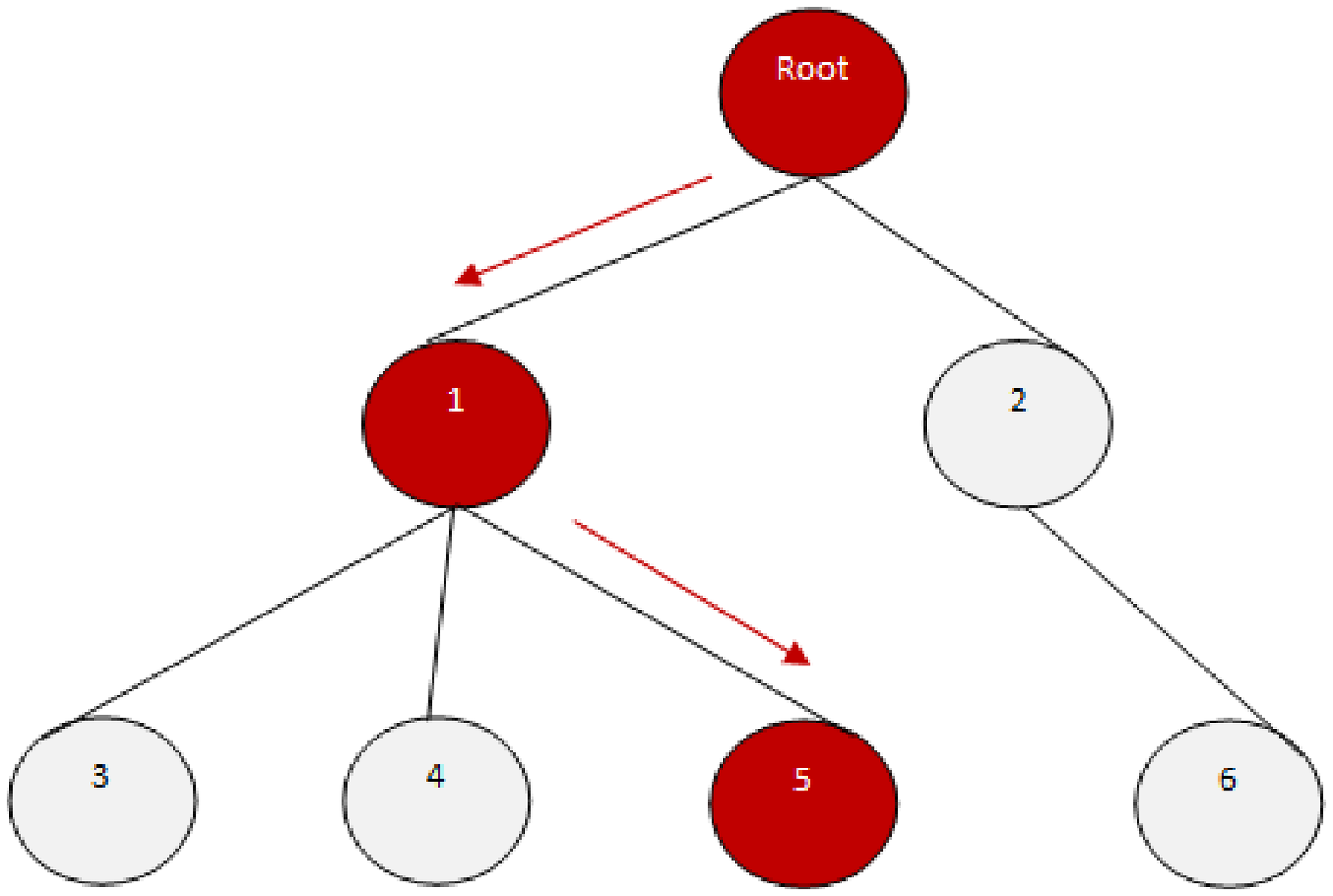}
\caption{Some relevant features of LCN methods: a case of inconsistency (on the left) and a typical top-down decision strategy in action (on the right).}
\label{fig:1}
\end{figure}


Being $\mathcal{T}$ the training set, $\Lambda_{i}$ examples whose most specific class is $c_{i}$, $T^{+}_{c_{i}}$ and $T^{-}_{c_{i}}$ positive and negative training set of $c_{i}$, there are many training policies as follows:

\begin{itemize}

\item \emph{Exclusive policy} \cite{Eisner}: $T^{+}_{c_{i}}=\Lambda_{i}$ and $T^{-}_{c_{i}}=T-T^{+}_{c_{i}}$ \\

\item \emph{Less exclusive policy} \cite{Eisner}: $T^{+}_{c_{i}}=\Lambda_{i}$ and $T^{-}_{c_{i}}=T-(\Lambda_{i} \cup \Downarrow \Lambda_{i})$ where $\Downarrow \Lambda_{i}$ is the set of descendent categories of $\Lambda_{i}$.\\

\item \emph{Less inclusive policy} \cite{Eisner}: $T^{+}_{c_{i}}=\Lambda_{i} \cup \Downarrow \Lambda_{i}$ and $T^{-}_{c_{i}}=T-T^{+}_{c_{i}}$\\

\item \emph{Inclusive policy} \cite{Eisner}: $T^{+}_{c_{i}}=\Lambda_{i} \cup \Downarrow \Lambda_{i}$ and $T^{-}_{c_{i}}=T-(\Lambda_{i} \cup \Downarrow \Lambda_{i} \cup \Uparrow \Lambda_{i})$ where $\Uparrow \Lambda_{i}$ is the set of ancestor categories of $\Lambda_{i}$.\\

\item \emph{Siblings policy} \cite{Sebastiani}: $T^{+}_{c_{i}}=\Lambda_{i} \cup \Downarrow \Lambda_{i}$ and $T^{-}_{c_{i}}= \Leftrightarrow {c_{i}} \cup \Downarrow (\Leftrightarrow {c_{i}})$ where $\Leftrightarrow {c_{i}}$ is the set of sibling categories of $\Lambda_{i}$.\\

\item \emph{Exclusive siblings policy} \cite{Ceci}: $T^{+}_{c_{i}}=\Lambda_{i}$ and $T^{-}_{c_{i}}=\Leftrightarrow {c_{i}}$ \\

\end {itemize}

The testing step can be performed in several ways. In the event that the output of each classifier is separately calculated for any incoming sample, this decision strategy is naturally multi-labeled. On the other hand,  class-membership inconsistency may occur.
To show a case of inconsistency, let us consider a case in which a sample belongs to nodes 1, 5, 2, and 6, while in fact the classifier that corresponds to node 2 has not been fired --see Figure~1 (left). This event, not so unlikely to occur, shows that some LCN methods are prone to class-membership inconsistency. Some methods have been devised to avoid inconsistencies, which force the selection of only one node at each level of the hierarchy \cite{Wu} \cite{Dumais} \cite{DeCoro}.


The top-down strategy is a commonly-used approach in LCN methods to avoid inconsistencies. This strategy assumes that the evaluation starts from the root and goes downward to the leaf --as shown in Figure~1 (right). At each level of the hierarchy, except for the root, the decision about which node to select at the current level is also based on the node predicted at the previous (parent) level. For example, suppose that the output of the local classifier for class 1 is true, and the output of the local classifier for class 2 is false. At the next level, the system will only consider the output of classifiers predicting classes which are children of class 1, i.e., nodes 3, 4 and 5.

Any top-down approach in which a stopping criterion permits the classification process to stop at any internal node of the underlying hierarchy
is prone to the so-called ``blocking problem'', which accurs when a classifier at a certain level in the class hierarchy predicts that the sample does not have the class associated with that classifier. In this case the sample will be ``blocked'', i.e., it will not be passed to the descendants of that node. This phenomenon happens whenever a threshold is used at each node, and if the confidence score or posterior probability of the classifier at a given
node (for a given test sample) is lower than this threshold, the classification disregards the incoming sample.

Moreover, top-down methods were originally forced to predict a leaf node, also known as mandatory leaf-node prediction in the literature. It is worth pointing out that a non mandatory leaf-node prediction setting, in combination with a top-down approach, does not prevent the blocking problem to occur, as the process can be stopped also due to an erroneous classification (false negative).

\section{The Proposed Label Evaluation Method for LCN}

In this section, we present the proposed label evaluation method for the LCN approach which enforces a top-down strategy for the testing phase. The proposed method tries to ensure the reliability of a candidate route in the hierarchy for a test sample before assigning the final label by the classifier. 
The idea is to identify the samples likely to be assigned the ``false'' label while they are in fact true. Once this is achieved, there are two options: to send the sample to another classification process or to simply reject the sample and send it to the manual labeling process. This decision is particularly crucial for the applications associated with a high cost of mislabeling true positives.

In the proposed method, we calculate the \emph{confidence score} for each selected node at each level of hierarchy as follows: 

\begin{equation}
CS(\hat{c})= \frac{\mathcal{P}(\hat{c})}{\sum_{\Leftrightarrow \hat{c}}{\mathcal{P}(\Leftrightarrow \hat{c})}} 
\end{equation}

\noindent where $\hat{c}$ is a node and $\mathcal{P}(c)$ is its posterior probability. This measure takes into account the confidence of the selected node compared to the rest of its siblings. 

Furthermore, in order to include the hierarchical information embedded in parent-child class relationships, weight factors are computed for each level of the hierarchy. These weights are calculated based on the accuracy of each level, so that a level with high error rate gets a reduced weight factor. While performing top-down evaluation of a sample, we calculate the \emph{reliability score} for the candidate route using formula 2.

Finally, using a threshold to decide about the label assigned to the candidate route generates an accept/reject answer or the application of another classifier designed for this purpose.

The threshold determined by Equal Error Rate (EER) leads to the equal false acceptance (FA) and false rejection (FR) rates. However, other strategies can also be considered. The proposed method is sketched in Algorithm 1 with more detail.

\begin{algorithm}[htbp]

\down

$\mathcal{T}r, \mathcal{T}v$ and $\mathcal{T}e$ are training, validation and testing sets, respectively. $\Theta$ is the classifier algorithm, which can be applied to any node of the hierarchy.\\

\textbf{Training}

\down

For each node of the hierarchy $c_{i}$ do:

\begin{enumerate}

\item Define the $T^{+}_{c_{i}}$ and $T^{-}_{c_{i}}$ from the $\mathcal{T}r$ set according to the policies given in Section~2.

\item Train a classifier $\hat{c_i} = \Theta(T^{+}_{c_i},T^{-}_{c_i})$

\end{enumerate}

\textbf{Validation}

\begin{enumerate}

\item For $d \in \mathcal{T}v$ apply $d$ to the all node classifiers

\item Calculate the recognition rate of each node on $\mathcal{T}v$ and use it as the weight factor for the node

\item For each $d$, calculate the reliability as follows:

\up

\begin{equation}
Reliability(d)= \sum_{level=1}^{L}{w(\hat{c}) \cdot CS(\hat{c})}
\end{equation}

\noindent where $w(\hat{c})$ and $CS(\hat{c})$ are the weight factors and confidence score of the selected node at each level of the hierarchy.

\item Plot histogram of the mislabeled vs. truly labeled reliability score for $\mathcal{T}v$ set, specify the Equal Error Rate (EER) where the false acceptance (FA) and false rejection (FR) are equal, and regard it as the threshold $\tau$ 

\end{enumerate}

\textbf{Testing}

\down

With $d \in \mathcal{T}e$:

\begin{enumerate}

\item Apply $d$ to the classifiers in top-down manner, starting from the root downward to reach the leaf

\item Calculate the reliability score using formula 2 for the candidate route.

\item If Reliability(d) $> \tau$ then accept the assigned label otherwise, reject it or follow another classification strategy.

\item Calculate the boosted accuracy as follows: 

\begin{equation}
Accuracy= \frac{number \hspace{1mm} of \hspace{1mm} truly \hspace{1mm} labeled \hspace{1mm} samples}{number \hspace{1mm} of \hspace{1mm} accepted \hspace{1mm} samples} 
\end{equation}

\end{enumerate}

\caption{Proposed label evaluation method for LCN. \label{MRPE}}
\end{algorithm}

\section{Experimental Results}

\highlight{Dataset description}

The Reuters Corpus Volume I (RCV1) \cite{Lewis} is a benchmark dataset widely used in text categorization and in document retrieval.
It consists of over 800,000 newswire stories, collected by the Reuters news and information agency. The stories have been manually coded using three orthogonal category sets. Category codes from three sets (Topics, Industries, and Regions) are assigned to stories:

\begin{itemize}

\item Topic codes capture the major subject of a story. The hierarchy of topics consists of a set of 104 categories organized in a four-level
hierarchy.

\item Industry codes are assigned on the basis of the types of business discussed in the story.

\item Region codes include both geographic locations and economic/political groupings.

\end{itemize}

We pre-processed documents proposed by Lewis et al. by retaining
only documents associated to a single category. This choice depends on 
the fact that in this study we are interested in investigating single category assignment (feature selection method, learning algorithms, categorization framework and performance evaluation functions are all based on the assumption that a document can be assigned to one category at the most).
We also separate the training set and the testing set using the same split adopted by Lewis et al.

\begin{table}
\centering
\caption{The main characteristics of the Reuter's RCV1-v2 datasets.}
\label{tab:1}       
\begin{tabular}{lllllllll}
\hline\noalign{\smallskip}
problem & \hspace{2mm} train & \hspace{2mm} test & \hspace{2mm} total samples & \hspace{2mm} classes & \hspace{2mm} levels &\hspace{2mm} class/L2 & \hspace{2mm} class/L3 & class/L4\\
\noalign{\smallskip}\hline\noalign{\smallskip}
topics & \hspace{2mm} 23,149 & \hspace{2mm} 781,265 & \hspace{2mm} 804,414 & \hspace{2mm} 104 & \hspace{2mm} 4 & \hspace{2mm} 4 & \hspace{2mm} 99 & \hspace{2mm} 1 \\
industry & \hspace{2mm} 23,149 & \hspace{2mm} 781,265 & \hspace{2mm} 804,414 & \hspace{2mm} 365 & \hspace{2mm} 4 & \hspace{2mm} 10 & \hspace{2mm} 354 & \hspace{2mm} 1 \\
regions & \hspace{2mm} 23,149 & \hspace{2mm} 781,265 & \hspace{2mm} 804,414 & \hspace{2mm} 366 & \hspace{2mm} 4 & \hspace{2mm} 7 & \hspace{2mm} 350 & \hspace{2mm} 9\\
\noalign{\smallskip}\hline
\end{tabular}
\end{table}

\highlight{Classifier description and experimental setup}

In TC applications the computational efficiency is crucial due to the very large number of features, classes, and samples size. Therefore, the issue of concerning the design of simple and fast classification systems is important. There are many research works in the literature using a variety of classifiers such as k-nearest neighbors (kNN), SVM, artificial neural networks, bayesian, and Rocchio's \cite{ML for TC}. However, in practice most of them are not applicable since in real-world applications (e.g., search engines, contextual advertising, recommender systems) the real-time requirement has great importance. Among them, the Rocchio classification algorithm is extremely simple and straightforward while showing competitive performance on text categorization problems. Moreover, it does not requires to store large amounts of training data. It calculates the prototype vector or centroid vector ($ C_{i} $) for class node ($c_{i}$):

\begin{equation}
C_{i}= \frac{1}{ \mid c_{i} \mid} \sum_{d \in c_{i}}{d}
\end{equation}

\noindent where $\mid A \mid$ denotes the cardinality of set $A$ and $d$ is the training document.

In the testing step, we calculate the similarity of one document $d$ to each centroid by the innerproduct measure,

\begin{equation}
S(d, C_{i})= d.C_{i}
\end{equation}

This similarity can be regarded as the \emph{posterior probability} of the node classifier and used for final decision making.

Moreover, to avoid the class-membership inconsistency problems, the node with max similarity have been selected at each level of the hierarchy. At the next level, the text sample have been applied only to the children of the selected node and so on till it reaches to the leaf. Therefore its also mandatory leaf-node prediction approach.\\

\highlight{Performance results}

In this subsection, we first show the performance of the proposed method in discriminating reliable vs. unreliable samples and then, rejecting the samples with the \emph{reliability score} lower and including the samples whose \emph{reliability score} is higher than the given threshold.
False rejections (FR) occur when the label is \emph{truly} assigned by the classifier while the reliability score is low or when the assigned label is false
while high reliability is given to the sample in hand.
Experimental results are reported in Table 2. As clearly shown, the proposed method rejects the samples falsely predicted by the classifier (TR), while the number of FR is very low when averaging on a large number of test samples.
It is clear that the number of FR and TR are directly related to the selected route evaluation threshold. In particular, higher thresholds reduce TR while increasing the FR. Hence, in applications with high cost for mislabeling, the proposed strategy can reduce the overall cost with the drawback of rejecting more truly-labeled samples.

\begin{table}
\centering
\caption{The results obtained by the proposed method on the Reuter's RCV1-v2 datasets.}
\label{tab:2}       
\begin{tabular}{lllll}
\hline\noalign{\smallskip}
problem & \hspace{2mm} rejected samples & \hspace{2mm} TR & \hspace{2mm} FR & \hspace{2mm} accuracy boost \\
\noalign{\smallskip}\hline\noalign{\smallskip}
topics & \hspace{2mm} 740 & \hspace{2mm} 652 & \hspace{2mm} 88 & \hspace{2mm} 8.2 \\
industry & \hspace{2mm} 602 & \hspace{2mm} 580 & \hspace{2mm} 22 & \hspace{2mm} 6.5  \\
regions & \hspace{2mm} 794 & \hspace{2mm} 598 & \hspace{2mm} 196 & \hspace{2mm} 7.8 \\
\noalign{\smallskip}\hline
\end{tabular}
\end{table}

For the sake of comparison, different widely acknowledged standard text categorization algorithms have been run and evaluated on the selected datasets. From these methods, the big-bang global method and flat are non-hierarchical while the LCP, LCL and LCN are hierarchical classification methods. To assess all the cited methods, a centroid-based classifier with the same parameters has been used. The results of this comparison is reported in Table 3, which clearly shows that the proposed method boosts the accuracy of the standard LCN method while outperforming both hierarchical and non-hierarchical methods.

\begin{table}
\centering
\caption{The proposed method outperforms the existing standard text categorization methods on the Reuter's RCV1-v2 datasets. (Recognition rate in percentage)}
\label{tab:3}       
\begin{tabular}{lllllll}
\hline\noalign{\smallskip}
problem & \hspace{2mm} big-bang & \hspace{2mm} flat & \hspace{2mm} LCP & \hspace{2mm} LCL & \hspace{2mm} LCN & \hspace{2mm} proposed method \\
\noalign{\smallskip}\hline\noalign{\smallskip}
topics & \hspace{2mm} 40.5 & \hspace{2mm} 41.2 & \hspace{2mm} 38.4 & \hspace{2mm} 38.9 & \hspace{2mm} 40.1 & \hspace{2mm} \bf{47.5} \\
industry & \hspace{2mm} 44.3 & \hspace{2mm} 43.0 & \hspace{2mm} 42.1 & \hspace{2mm} 41.3 & \hspace{2mm} 42.7 & \hspace{2mm} \bf{47.4} \\
regions & \hspace{2mm} 44.0 & \hspace{2mm} 44.5 & \hspace{2mm} 42.5 & \hspace{2mm} 42.8 & \hspace{2mm} 43.0 & \hspace{2mm} \bf{48.9} \\
\noalign{\smallskip}\hline
\end{tabular}
\end{table}

\section{Conclusions and Future Work}

A route confidence evaluation method is proposed for reliable HTC. The main strength of the proposed method concerns the integration of a prediction mechanism able to identify and deal with the samples which would be wrongly labelled by the classifier. This clearly results in a boost in accuracy of the LCN method by simply rejecting a low percentage of the test samples.

Experimental results on the Reuters RCV1-v2 datasets show a significant improvement in the recognition rate, compared to the standard LCN method. Furthermore, a comparison with results obtained by running other TC methods emphasizes an overall superior performance of the proposed method.

\subsubsection*{Acknowledgments.} 
Camelia Chira acknowledges the support of Grant PN II TE 320, Emergence, auto-organization and evolution: New computational models in the study of complex systems, funded by CNCS Romania.

\end{document}